\documentclass{article}

\usepackage[final]{neurips_2023_ml4ps}

\usepackage[utf8]{inputenc} 
\usepackage[T1]{fontenc}    
\usepackage{hyperref}       
\usepackage{url}            
\usepackage{booktabs}       
\usepackage{amsfonts}       
\usepackage{nicefrac}       
\usepackage{microtype}      
\usepackage{xcolor}         
\usepackage{graphicx}
\usepackage{subcaption}

\newcommand{\fink}{{\sc Fink}}

\usepackage{xcolor,soul}

\title{Bayesian multi-band fitting of alerts for kilonovae detection}

\author{
  Biswajit Biswas \\
  AstroParticule et Cosmologie\\
  Université Paris Cité, CNRS\\
  F-75013, Paris, France \\
  \texttt{biswas@apc.in2p3.fr} \\
  \And
  Junpeng Lao \\
  Google Switzerland \\
  Gustav-Gull-Platz 1 \\
  8004 Zurich, Switzerland \\
  \AND
  Eric Aubourg \\
  AstroParticule et Cosmologie \\
  Université Paris Cité, CNRS\\
  F-75013, Paris, France \\
  \And
  Alexandre Boucaud \\
  AstroParticule et Cosmologie \\
  Université Paris Cité, CNRS\\
  F-75013, Paris, France \\
  \And
  Axel Guinot \\
  AstroParticule et Cosmologie \\
  Université Paris Cité, CNRS\\
  F-75013, Paris, France \\
  \And
  Emille E. O. Ishida \\
  LPC, Université Clermont Auvergne \\
  CNRS/IN2P3\\
  F-63000, Clermont-Ferrand, France \\
  \And
  Cécile Roucelle \\
  AstroParticule et Cosmologie \\
  Université Paris Cité, CNRS\\
  F-75013, Paris, France \\
}

\begin{document}

\maketitle

\begin{abstract}
In the era of multi-messenger astronomy, early classification of photometric alerts from wide-field and high-cadence surveys is a necessity to trigger spectroscopic follow-ups. These classifications are expected to play a key role in identifying potential candidates that might have a corresponding gravitational wave (GW) signature. 
Machine learning classifiers using features from parametric fitting of light curves are widely deployed by broker software to analyze millions of alerts, but most of these algorithms require as many points in the filter as the number of parameters to produce the fit, which increases the chances of missing a short transient.
Moreover, the classifiers are not able to account for the uncertainty in the fits when producing the final score.
In this context, we present a novel classification strategy that incorporates data-driven priors for extracting a joint posterior distribution of fit parameters and hence obtaining a distribution of classification scores.
We train and test a classifier to identify kilonovae events which originate from binary neutron star mergers or neutron star black hole mergers, among simulations for the Zwicky Transient Facility observations with 19 other non-kilonovae-type events. 
We demonstrate that our method can estimate the uncertainty of misclassification, and the mean of the distribution of classification scores as point estimate obtains an AUC score of 0.96 on simulated data. 
We further show that using this method we can process the entire alert steam in real-time and bring down the sample of probable events to a scale where they can be analyzed by domain experts.
\end{abstract}

\section{Introduction}

The detection of the binary neutron star (BNS) merger event GW170817 \citep{PhysRevLett.119.161101} marked the dawn of multi-messenger astronomy as the electromagnetic (EM) counterpart of the event was detected following the initial GW detection by LIGO and VIRGO collaborations. 
This EM counterpart, so-called kilonova (KN), originates from BNS and possibly from neutron star black hole mergers and is a source of heavy element production via the r-process\citep{2017ApJ...851L..21V}. 
Moreover, detecting such events along with the GW counterpart gives us a way to evaluate the value of Hubble's Constant ($H_0$) \citep{1986Natur.323..310S} and addresses the tension between the Cosmic Microwave Background and Type Ia supernovae (SNIa) measurements of $H_0$\citep{2021ApJ...919...16F}.

To a certain extent, GW detectors can localize events in the sky. 
However, a large sky fraction needs to be scanned to look for EM counterparts within the localization error boxes obtained from GW signal. 
Although spectroscopic measurement is necessary to obtain an accurate spectrum of these events, wide-field surveys observe a much bigger patch of the sky in broadband filters so they can be used to select targets for deployment of the expensive spectroscopic follow-up. 
Hence considerable efforts are being put into identifying potential events using the Zwicky Transient Facility \cite[ZTF,][]{Bellm_2019, Andreoni:2021},  Dark Energy Camera\citep{Flaugher_2015, Garcia2020} and also preparing for the arrival of data from the Rubin Observatory Legacy Survey of Space and Time \cite[LSST,][]{2019ApJ...873..111I}.

Whenever a $5\sigma$ signal is detected above the background, ZTF creates alerts containing a 30-day history of the event.
Since around one million such alerts are generated each night by ZTF, intermediary broker software is required to process the incoming alert stream and identify events of interest. 
Typical methods of classification rely on parameters obtained from performing parametric fits of the light curves, independently on each filter. 
A major challenge in identifying KNe events in the alert stream is the limited availability of photometric detection points in each broadband filter. 
As we move from ZTF to LSST, we will have more photometric bands, but fewer points per band.  Hence we need to prepare for the complexity of classification with scarce data.

In the context of the \fink\ broker \citep{10.1093/mnras/staa3602} that is being developed for LSST and is currently processing alerts from the ZTF, we present a new approach to combine information from multiple bands and jointly fit all the filters of an alert by performing a Bayesian fit using Monte Carlo Markov Chain (MCMC) with a data-driven prior. 
By performing a simultaneous fit, we remove the constraint of a minimum number of points and also provide a distribution of the scores of the classification.
Although our approach can be extended to any parametric fitting, in this work we use the parametric fitting scheme presented in \citet{refId0}.

\section{Dataset}

To train and test our classifier we use simulations from \citet{Muthukrishna_2019} generated with SNANA simulator \citep{Kessler_2009} and models from the Photometric LSST Astronomical Classification Challenge
\cite[PLAsTiCC,][]{Kessler2019}.
This set consists of 42568 events among which 38000 are
non-KNe representing 19 other types of events, while the remaining 4568 KNe simulations comprise 2 KN models (see \citet{Muthukrishna_2019} for details). 
Moreover, to represent a more recent ZTF cadence, another 1000 KNe events from \citet{10.1093/mnras/staa1776} were added to this dataset.

To mimic alert data, we consider 30 days of data around the observation with the highest signal-to-noise ratio, starting 10 days before and running up to 20 days after.
This assumption is not exactly true for alerts because we will have information only up to the day of the alert.
However, it serves as a necessary first test to evaluate our method before testing on a more complex setting.
This combined dataset is then split into a training set of 3280 KNe and 19000 non-KNe and a test set of 2288 KNe and 19000 non-KNe. For more details, readers can refer to \citet{refId0}. 

\section{Methodology}

\begin{figure}
     \centering
     \begin{subfigure}[t]{0.45\textwidth}
         \includegraphics[width=\textwidth]{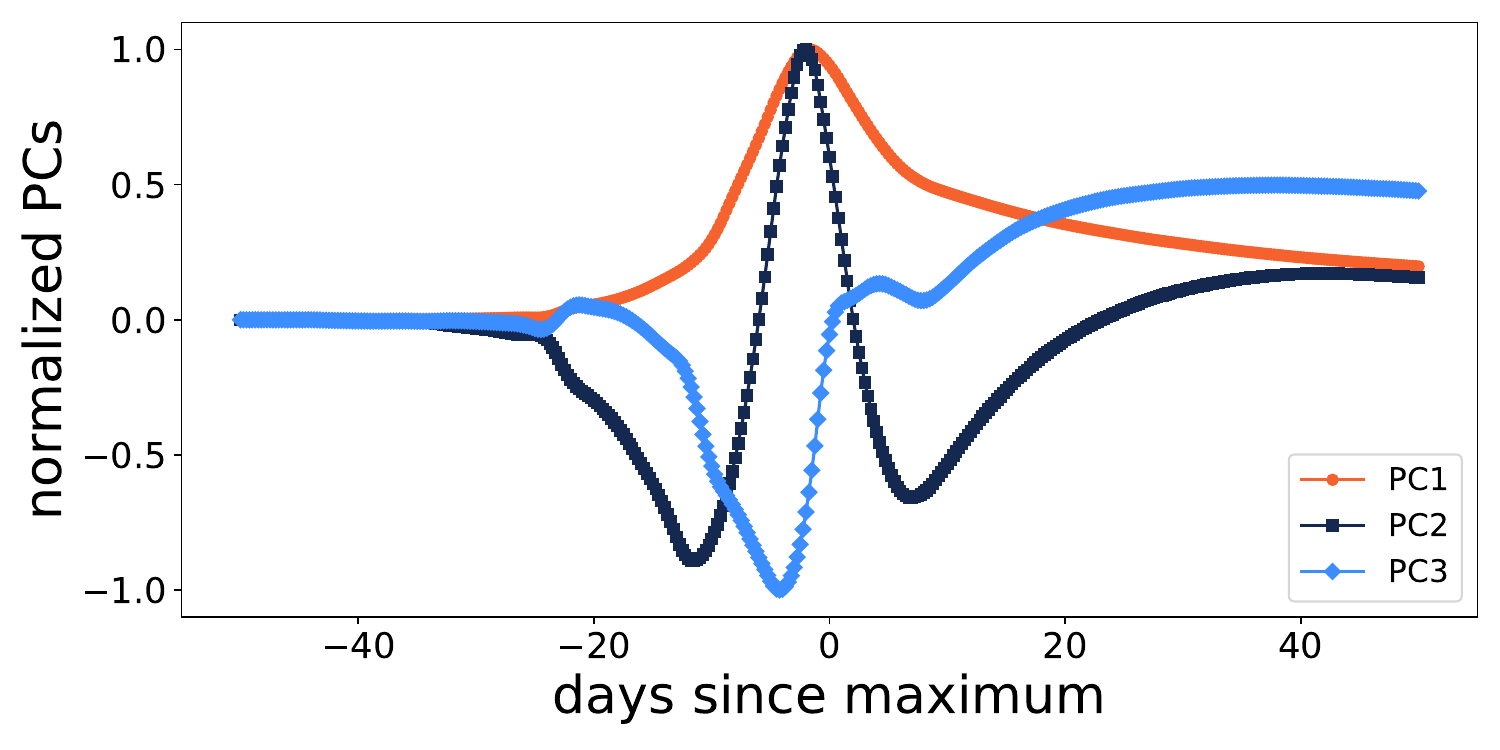}
         \caption{\centering Basis vectors for reconstructing light curves where each principal component is used as a template in the linear combination for generating fits.}
         \label{fig: PCS}
     \end{subfigure}
     \hspace{0.025\textwidth}
     \begin{subfigure}[t]{.45\textwidth}
         \includegraphics[width=\textwidth]{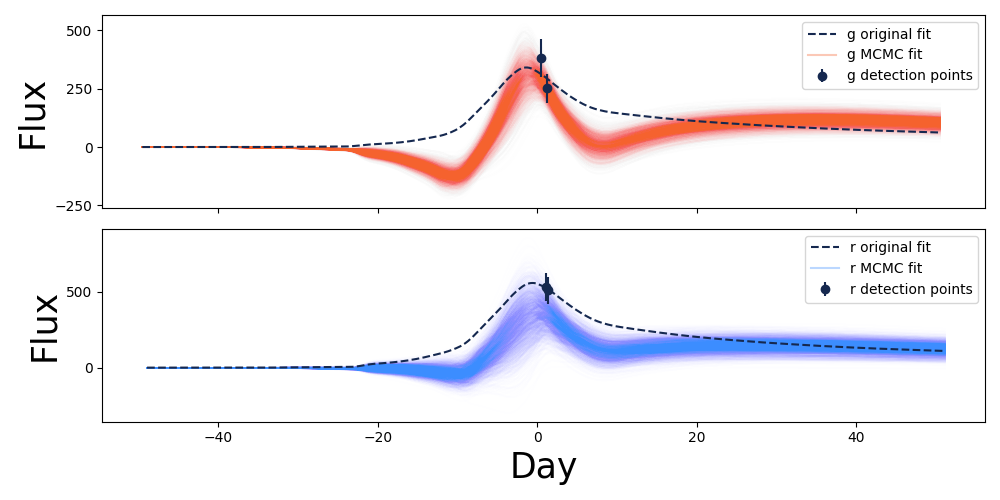}
         \caption{\centering Examples of reconstructions using basis vectors where dashed black lines are generated from the original work, and colored lines represent samples from our fit.}
         \label{fig: reconst example}
     \end{subfigure}
     \caption{\centering Parametric fitting of light curves.}
     \label{fig: parametric fit}
\end{figure}

To demonstrate the performance of our approach we use the method proposed by \citet{refId0} where each filter of the light curve is represented as a linear combination of three basis vectors generated by performing principal component analysis on noiseless light curve realizations. 
Figure \ref{fig: PCS} shows the set of templates that are used as basis vectors to fit the light curves from day $-50$ to $50$, where day $0$ is approximately the maximum observed flux.
The optimized coefficients of this linear combination constitute a set of features that are fed to a random forest \cite[RF,][]{ho1995random} classifier to classify alerts as KNe by the authors. 
A limitation of such approaches is that the parametric fit is performed independently on each filter. It requires at least as many points in each filter as the number of free parameters. 
Moreover, the uncertainty of the fit is not taken into account during the classification process since the fit is deterministic.

In this work, we first develop a data-driven prior for the joint distribution of parameters of all the filters. 
We combine the prior with Gaussian log-likelihood for the observed points to generate MCMC chains to obtain a posterior distribution for the parameters. 
Figure \ref{fig: reconst example} shows an example of the distribution of fits we obtained from our algorithm instead of the deterministic fit proposed by \citep{refId0}. 
The authors generate fits even when there are fewer points than the number of free parameters by using strong regularization, but they still require at least $2$ points in each band.
As we move towards LSST, such a constraint on all the $6$ LSST bands will result in missing faster transients like KNe. Hence, motivating a joint multi-band Bayesian fitting to eliminate this constraint.

\subsection{Generating features}

To build the data-driven prior, we need to identify a sample of events for which we are confident that we have good fits.
Ideally, the goal would be to construct this prior from real data but due to the lack of availability to real KNe events, we rely on the features generated by \citet{refId0} on the realistic training set simulations.
These features were generated only on events with multiple points in each filter thus ensuring reliable fits.
As KNe events are short and more difficult to fit, we extract a joint multi-band covariance matrix from the features of only the KNe events to act as a prior for our Bayesian fit.

Although we have all the necessary components, before running the MCMC chains we must normalize the light curve data in the same manner as \citet{refId0}. 
We thus divide the data of each band by the maximum observed flux in that band. 
Once the parameters of the fit are obtained, we include this information, once again, in the classifier by passing the normalization factor as a separate feature.
For each event, after a warmup phase of 500 steps, we generate 4 MCMC chains with 500 samples in each chain which provides us with a posterior distribution of 2000 samples per event.

\subsection{Classification}

One approach for classification could be to use summary statistics such as the mean and standard deviation of the posterior distribution to obtain a discrete score for the event. 
But in a high-dimension space, if the distribution is non-Gaussian, these summary statistics will not encompass all the information. 
Instead, we chose to obtain a distribution of classification scores by classifying each posterior sample individually.

\begin{figure}[t]
   
   \begin{minipage}{0.45\textwidth}
     \centering
     \includegraphics[width=\linewidth]{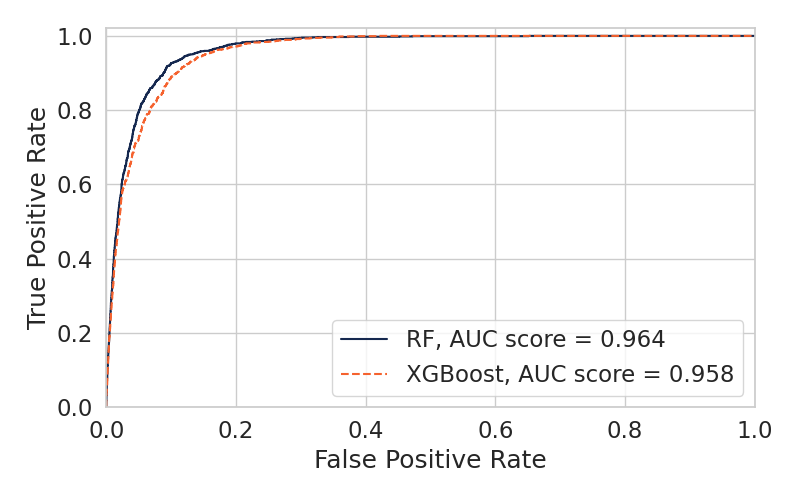}
     \caption{\centering Receiver operating characteristics curve.}
     \label{fig: roc curve}
   \end{minipage}\hfill
   \begin{minipage}{0.45\textwidth}
     \centering
     \includegraphics[width=\linewidth]{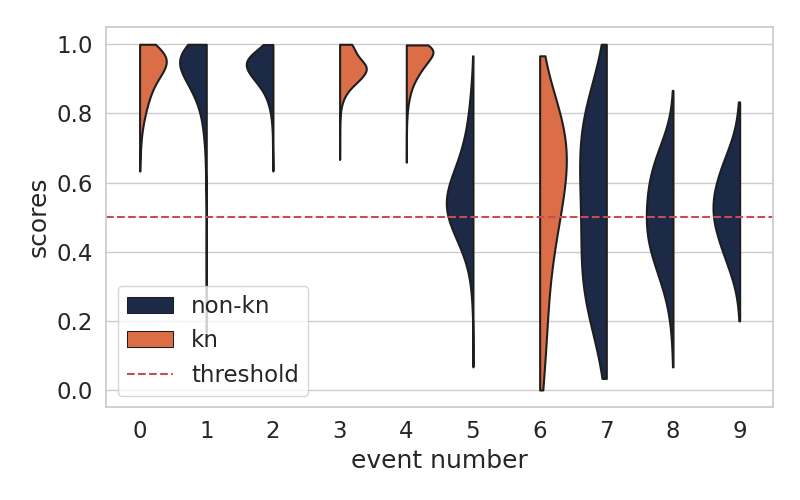}
     \caption{\centering Examples of distribution of scores for $5$ events with mean score close to $1$ and $5$ events near the classification threshold of $0.5$.}\label{fig: scores distrib}
   \end{minipage}
\end{figure}

Along with the fit parameters, we also pass as features to the classifier the maximum value of flux in each band and the log-likelihood of the fit. 
We compared an RF using \texttt{scikit-learn} \citep{scikit-learn} and XGBoost classifier \citep{Chen:2016:XST:2939672.2939785} with 30 trees and a max depth of 42 as no significant performance gain was observed with a more complicated model. For training, we used a balanced set of 200k samples drawn from the complete pool of samples and found that the performances were comparable(Figure \ref{fig: roc curve}). 
Hence, we hereafter consider only the results from the RF classifier.
Figure \ref{fig: scores distrib} shows the distribution of scores for some events with mean scores around $0.5$ and some around $1$. 
As we do not observe a bimodal distribution for the scores, we use the mean and standard deviation of the distribution for each alert as summary statistics to draw our final conclusions.

\section{Results}

Upon using the classifier on the test sample, we obtain the ROC curve shown in figure \ref{fig: roc curve} using the mean as a point estimate of the classification score. 
The exact choice of threshold would depend upon the application of such a classifier.
For example, as there are around one million alerts from ZTF every night, we might want to restrict ourselves to a smaller false positive (FP) rate so that we have a reliable sample that can be followed up by experts. 
On the other hand, if we have a GW localization in a particular region of the sky, our goal would be to not miss the event, even if it is at the cost of a high FP rate.

\begin{figure}
     \centering
     \begin{subfigure}[t]{0.4\textwidth}
         \includegraphics[width=\textwidth]{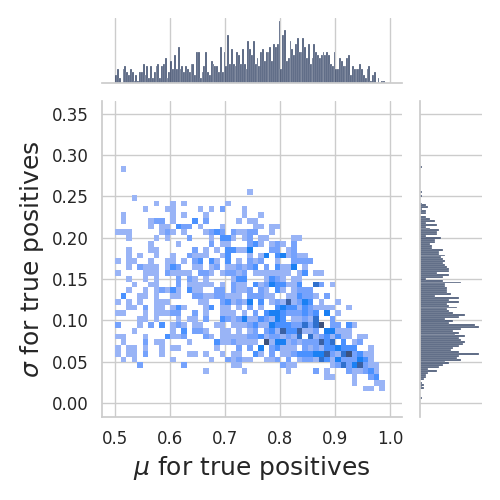}
         \caption{\centering Distribution of scores and sigma for True positives.}

     \end{subfigure}
     \hspace{0.05\textwidth}
     \begin{subfigure}[t]{.4\textwidth}
         \includegraphics[width=\textwidth]{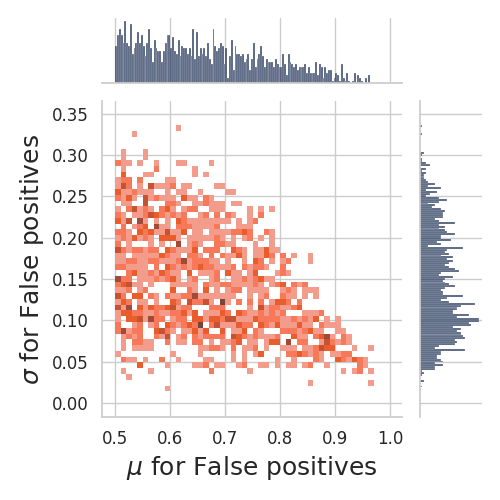}
         \caption{\centering Distribution of scores and sigma for False positives.}

     \end{subfigure}
     \caption{Distribution of mean($\mu$) and sigma($\sigma$) for each event in test set.}
     \label{fig: mu sig distrib}
\end{figure}

Since the classification threshold should ideally be dynamic, we present results with the threshold set to $0.5$. 
Under this setting, we obtain a precision of $0.45$ and recall of $0.85$.
Even though the value of precision might seem fairly low, we can bring down the number of potential events to a scale where they can be analyzed by domain experts while maintaining a high value for the recall, i.e., without missing many actual KNe events. 
In Figure \ref{fig: mu sig distrib} we further observe that the FP events have relatively lower values of scores and a higher value of uncertainty associated with the classification hence validating the uncertainty quantification.
A direct comparison with the results of \cite{refId0} where the authors require at least 2 points in each band, we obtain a drop in performance in terms of both precision and recall. 
However, such a comparison can be misleading as our sample contains events with less than 2 data points per band as well since there is no constraint on the minimum number of points.

\section{Conclusion and Outlook}

As we prepare for the arrival of LSST, we need robust classification algorithms that can operate on a small number of points for early classification of short-term transients. 
Our Bayesian fitting approach for KN classification takes a step in this direction by jointly fitting all filters, thereby combining information from all photometric bands to generate a parametric fit.
Our algorithm is not only able to incorporate information from a data-driven prior but also estimate the uncertainty of the fits. 

As the next step, we must retrain the classifier on more alert-like curves with no data after the day of the alert.
Although a slight drop in performance is expected because there will be incomplete data about the light curve, we expect the results to generalize fairly well.
Once trained, it is straightforward to integrate the method proposed here within the \fink\ broker to be tested on real ZTF alerts before being extended for the 6 bands of LSST.
Additionally, the computational cost needs to be taken into account when pushing the module into production.
We were able to generate features for around 1000 events per minute using a single CPU. The algorithm being written in the JAX\citep{jax2018github} framework, can easily be parallelized for faster computing efficiency. 
Although our algorithm is capable of processing the entire ZTF alert stream, processing the LSST stream, which will be at least one order of magnitude larger, would still be a challenge.
However, we can easily introduce appropriate selections to discard a major fraction of events at an earlier stage to significantly reduce the computational cost.
For LSST, in particular, this module will be used in combination with a series of selection cuts which would result in a feasible mechanism to analyze the alert stream and filter out interesting events in real-time so that spectroscopic follow-up can be triggered on these events.

\begin{ack}
We deeply thank Daniel Muthukrishna and Cosmin Stachie for providing a set of simulated KNe data used to train the model. 
This project has received funding from the European Union’s Horizon 2020 research and innovation program under the Marie Skłodowska-Curie grant agreement No 945304 – Cofund AI4theSciences hosted by PSL University.
This work was developed within the \fink\ community and made use of the \fink\ community broker resources. \fink\ is supported by LSST-France and CNRS/IN2P3. 

\end{ack}

\section*{References}

\bibliography{refs}

\end{document}